\documentclass[reprint,amsmath,amssymb,aps,prb,amssymb,superscriptaddress]{revtex4-2}
\usepackage[utf8]{inputenc}
\usepackage{color}
\usepackage{amsmath}
\usepackage{mathtools}
\usepackage{bm}
\usepackage{graphicx}
\usepackage{ulem}
\usepackage{amsfonts}
\usepackage{dcolumn}
\usepackage{natbib}
\usepackage{bbold}
\usepackage{soul}
\usepackage[dvipsnames]{xcolor}
\usepackage{amssymb}
\usepackage{mathrsfs}
\usepackage{rsfso}
\usepackage{bm}

\usepackage[breaklinks=true,colorlinks,citecolor=blue,linkcolor=blue,urlcolor=blue]{hyperref}
\def\be{\begin{equation}}
\def\ee{\end{equation}}
\def \bea{\begin{eqnarray}}
\def \eea{\end{eqnarray}}

\begin{document}
\title{Strain induced phase transition from antiferromagnet to altermagnet}
\author{Atasi Chakraborty}
%\email{atasi.chakraborty@uni-mainz.de}
\affiliation{Institut f\"{u}r Physik, Johannes Gutenberg Universit\"{a}t Mainz, D-55099 Mainz, Germany}

\author{Rafael Gonz\'{a}lez Hern\'{a}ndez}
\affiliation{Departamento de Fisica y Geociencias, Universidad del Norte, Barranquilla 080020, Colombia}
\author{Libor \v{S}mejkal}
\affiliation{Institut f\"{u}r Physik, Johannes Gutenberg Universit\"{a}t Mainz, D-55099 Mainz, Germany}

\affiliation{Institute of Physics, Academy of Sciences of the Czech Republic, Cukrovarnick\'{a} 10, 162 00 Praha 6, Czech Republic}
\author{Jairo Sinova}
\email{sinova@uni-mainz.de}
\affiliation{Institut f\"{u}r Physik, Johannes Gutenberg Universit\"{a}t Mainz, D-55099 Mainz, Germany}
\affiliation {Department of Physics, Texas A \& M University, College Station, Texas 77843-4242, USA}
%\affiliation{Institute of Physics, Academy of Sciences of the Czech Republic, Cukrovarnick\'{a} 10, 162 00 Praha 6, Czech Republic}

%\date{\today}

%%%%%%%%%%%%%%%%%%%%%%%%%% Guide lines %%%%%%%%%%%%%%%%%%%%%%%%%%%%%%
% Nature articles guide line of word count: the main text (not including Abstract, Methods, References and Figure legends) should be limited to 5,000 words

% Nature summary/ abstract paragraph guideline of journal: 
% - One or two sentences providing a basic introduction to the field, comprehensible to a scientist in any discipline. 
% - Two to three sentences of more detailed background, comprehensible to scientists in related disciplines.
% - One sentence clearly stating the general problem being addressed by this particular study.
% - One sentence summarizing the main result (with the words “here we show” or their equivalent).
% - Two or three sentences explaining what the main result reveals in direct comparison to what was thought to be the case previously, or how the main result adds to previous knowledge.
% - One or two sentences to put the results into a more general context.
% - Two or three sentences to provide a broader perspective, readily comprehensible to a scientist in any discipline, may be included in the first paragraph if the editor considers that the accessibility of the paper is significantly enhanced by their inclusion. Under these circumstances, the length of the paragraph can be up to 300 words. (This example is 190 words without the final section, and 250 words with it).
%%%%%%%%%%%%%%%%%%%%%%%%%%%%%%%%%%%%%%%%%%%%%%%%%%%%%%%%%%%%%%%%%%%%%%

\begin{abstract}
The newly discovered altermagnets are unconventional collinear compensated magnetic systems, exhibiting even (d, g, or i-wave) spin-polarization order in the band structure, setting them apart from conventional collinear ferromagnets and antiferromagnets. Altermagnets offer advantages of spin polarized current akin to ferromagnets, and THz functionalities similar to antifferomagnets, while introducing new novel effects like spin-splitter currents. A key challenge for future applications and functionalization of altermagnets, is to demonstrate controlled transitioning to the altermagnetic phase from other conventional phases in a single material. Here we prove a viable path towards overcoming this challenge through a strain-induced transition from an antiferromagnetic  to an altermagnetic phase in  ReO$_2$. Combining spin group symmetry analysis and \textit{ab-initio} calculations, we demonstrate that  under compressive strain ReO$_2$ undergoes such transition, lifting the Kramer's degeneracy of the band structure of the antiferromagnetic phase in the non-relativistic regime. In addition, we show that this magnetic transition is accompanied by a metal insulator transition,
%change in the nontrivial surface state topology from one phase to the other. 
and calculate the distinct spin polarized spectral functions of the two phases, which can be detected in angle resolved photo-emission spectroscopy experiments.
\end{abstract}
\maketitle

\section{Introduction}

%\textit{first para: generic introduction of altermagnetism}\\
The  recently discovered altermagnets (AMs) are a new \textit{third} class of collinear compensated magnetic materials, 
~\cite{Smejkal2020, Smejkal2021a, Smejkal2022a, Mazin2021, Mazin2022a, Smejkal2022GMR, Jaeschke-Ubiergo2023} that go beyond the conventional collinear ferromagnetic and antiferromagnetic classes.
AMs are  compensated magnetic ordered systems exhibiting  unconventional spin-polarized d/g/ or i-wave order in the non-relativistic 
band structure, originating from local sublattice symmetries %anisotropies 
in direct space. This gives rise to properties unique to altermagnets (e.g., the spin-splitter effect), while also giving similar properties of ferromagnets (FMs) (e.g., polarized currents) and antiferromagnets (AFMs) (e.g., THz spin dynamics and zero net magnetization). The unconventional time reversal symmetry (TRS) breaking in momentum space in AMs arises from the spin point-group symmetries \cite{Litvin1974,Smejkal2021a} that interchange sublattices with opposite spins only via a rotation (proper or improper).

The merged advantageous technological characteristics of AMs,  
makes controlling the transition between AMs and FMs/AFMs, not only of fundamental interest, but also serves as a foundational element for device applications. Tunable AMs can be used in data storage, sensors, and spintronics applications by exploiting, e.g., efficient spin current generation~\cite{Shao2021, Gonzalez-Hernandez2021, Bose2022}, giant magneto-resistance~\cite{Smejkal2022GMR}, spin-splitter torques~\cite{Bai2021, Karube2022}, anomalous Hall effects~\cite{Smejkal2020, Feng2022, Betancourt2021, Guo2023}, and Josephson effects~\cite{Oussou2023}. Because of the intricate connection of crystal symmetry and altermagnetism, mechanical deformation offers the most direct route to modify electronic properties solely by altering crystal symmetries, without necessitating additional electronic perturbations or chemical modifications.
Tensile strain and hydrostatic pressure are established reliable techniques to directly control lattice parameters and crystal symmetry, and hence can  control phases connected with magnetism, superconductivity, and topology~\cite{Chu2010, Hicks2014, Mutch2019, Xu2022, Li2017}. 

In this article, we propose to use %present 
strain as a controlling parameter to tune magnetic transition  from a conventional AFM to an AM. We choose  ReO$_2$ as working platform to explore phase transitions with mechanical deformations, because of its non-trivial topology emerging from non-symmorphic symmetries \cite{Wang2017, Hirai2021} and its recently observed structural phases \cite{Shibata2020, Ivanovskii2005}. Bulk ReO$_2$ primarily crystallizes in monoclinic $\alpha$- and orthorhombic $\beta$- phases. The topological characteristics of orthorhombic $\beta$-ReO$_2$ %with space group $Pbcn$ 
are of particular interest due to its hour glass like Dirac chains~\cite{Wang2017} and proposed to order antiferromagnetically around $\sim$4.2 K, while the monoclinic phase of ReO$_2$ is less explored. However, a recent experimental report has revealed a transformation from the monoclinic phase into the tetragonal rutile ($R$) phase driven by compressive lattice strain with decreasing film thickness~\cite{Shibata2020}. The spin group theory utilized to classify and characterize altermagnets \cite{Smejkal2021a}, predicts $R$-ReO$_2$ to be a $d$-wave AM candidate.
Although there are no reports on magnetic transitions in the magnetic system class of ReO$_2$, the hydrostatic pressure-induced continuous `martensitic' transition between the monoclinic to the tetragonal phase in the context of other transitions, e.g. the metal insulator transitions (MIT), for the ZrO$_2$ systems, to which ReO$_2$ belongs, has been well studied in the literature~\cite{Bailey1964, Simeone2003, Asgerov2022}.

Using both symmetry analysis and state of the art density functional theory (DFT) calculations, we show that the pressure-induced deformation, and their corresponding changes in the crystal symmetry, lead to an AFM to AM crossover in ReO$_2$. Further we investigate the strain induced change in topology and nodal crossing within the same chemical composition of ReO$_2$. As the AFM to AM transition is insensitive to susceptibility measurements, we provide a way to experimentally sense the phase transition through spin-polarized angle resolved photo emission spectroscopy (S-ARPES) experiment~\cite{Damascelli2004,Fedchenko2023}. 

\section{Results}
\subsection{Strain mediated phase transition}
In this section, we provide the structural comparison between the usual monoclinic ($\alpha-$) phase and the non-trivial tetragonal ($R-$) phase of ReO$_2$, %{\red and delve further into the intricate realm of} 
and explore the structural transition. %with change in volume. 
\begin{figure}[b]
\begin{center}
 \includegraphics[width=\columnwidth]{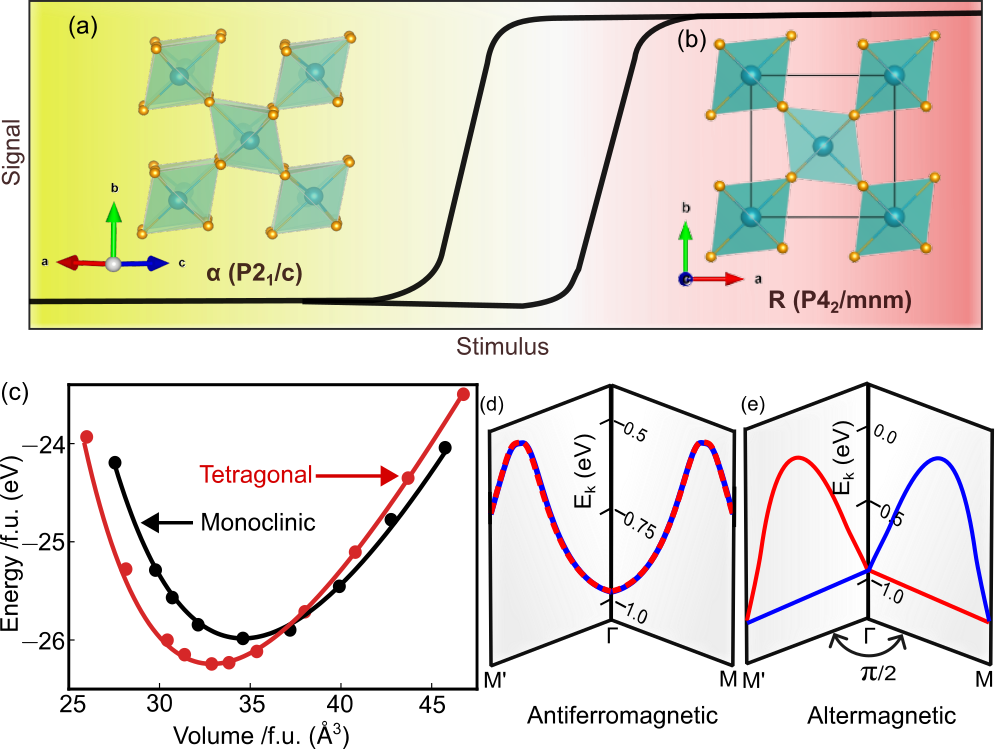}
 \caption{\textbf{Phase transition:} The top panel depicts a schematic representation of the martensitic structural transition from a monoclinic phase to a tetragonal phase under the influence of external stimuli. Inset (a) and (b) shows the u.c. of  monoclinic $\alpha-$ReO$_2$ and tetragonal $R-$ReO$_2$ respectively. (c) Birch Murnaghan fitting of DFT cohesive energy vs volume of the two ReO$_2$ phases. The presence of a stable tetragonal phase under compressive strain is in agreement with experimental observations. Spin polarized energy dispersions of antiferromagnetic (AFM) monoclinic (d) and altermagnetic (AM) tetragonal (e) ReO$_2$ along orthogonal momentum paths.  
  \label{fig_sch}}
 \end{center}
\end{figure}
Bulk $\alpha$-ReO$_2$ crystallizes in monoclinic space group $P2_{1}/ c$ (14). The distorted ReO$_6$ octahedra accommodating the nearest neighbor Re-O bonds ($\sim 1.86$~\AA$ -2.13$~\AA) form a geometrical network which is edge sharing in one direction and corner shared along the other. A recent experimental study validates the presence of tetragonal $R$-ReO$_2$ induced by strain, which exhibits the space group $P4_2{\slash}mnm$ \cite{Shibata2020}. Here, we will calculate the transition pathways between the two phases. We summarize the local structural parameters for both $R$- and $\alpha$-ReO$_2$ in Table~I. 
\begin{table}[h!]
\caption{Comparison of the tetragonal and monoclinic lattice parameters. 
%The lattice constants of tetragonal structure are a/b= 4.95~\AA, c=2.68~\AA, whereas that of monoclinic structure are a=5.50~\AA, b=4.81~\AA and c=5.58~\AA. 
In the monoclinic unit cell each ReO$_6$ octahedra contains six different bond lengths of octahedra. Minimum and maximum Re-O bond lengths are quoted here~$^{||}$.}
\label{Tab1}
\begin{tabular}{l c c} 
 \hline \hline
 &~ {Tetragonal}~ & ~{Monoclinic}~ ~ \\
 \hline 
$\beta$ & 90$^o$  & 119.55$^o$ \\
Re-O (\AA) & 1.96,~2.04 & 1.86-2.13$^{||}$ \\
Re-Re [edge] (\AA)  & 2.68 & 3.07 \\
Re-Re [corner] (\AA) & 3.75 & 3.55, 3.61, 3.81 \\
\hline \hline
\end{tabular}
\end{table}
Our calculations show that the %presence of first-order 
$\alpha$- to $R$-phase transition is accompanied by %nearly 
a $\sim$2.5\% volume collapse per formula unit. The unit cell (u.c.) structure along with the schematic of `martensitic' kind transition is shown in the top panel of Fig.~\ref{fig_sch}. The u.c. of $\alpha$- and $R$- ReO$_2$ are plotted in Fig.~\ref{fig_sch} (a) and (b) respectively.
%shown in Fig.~\ref{fig_sch}~(a) {\red and (b)?}. 

%%%%%%%%%% Figure 2 %%%%%%%%%%%%%%%%%%%%
\begin{figure*}[t!]
\begin{center}
 \includegraphics[width=1.75\columnwidth]{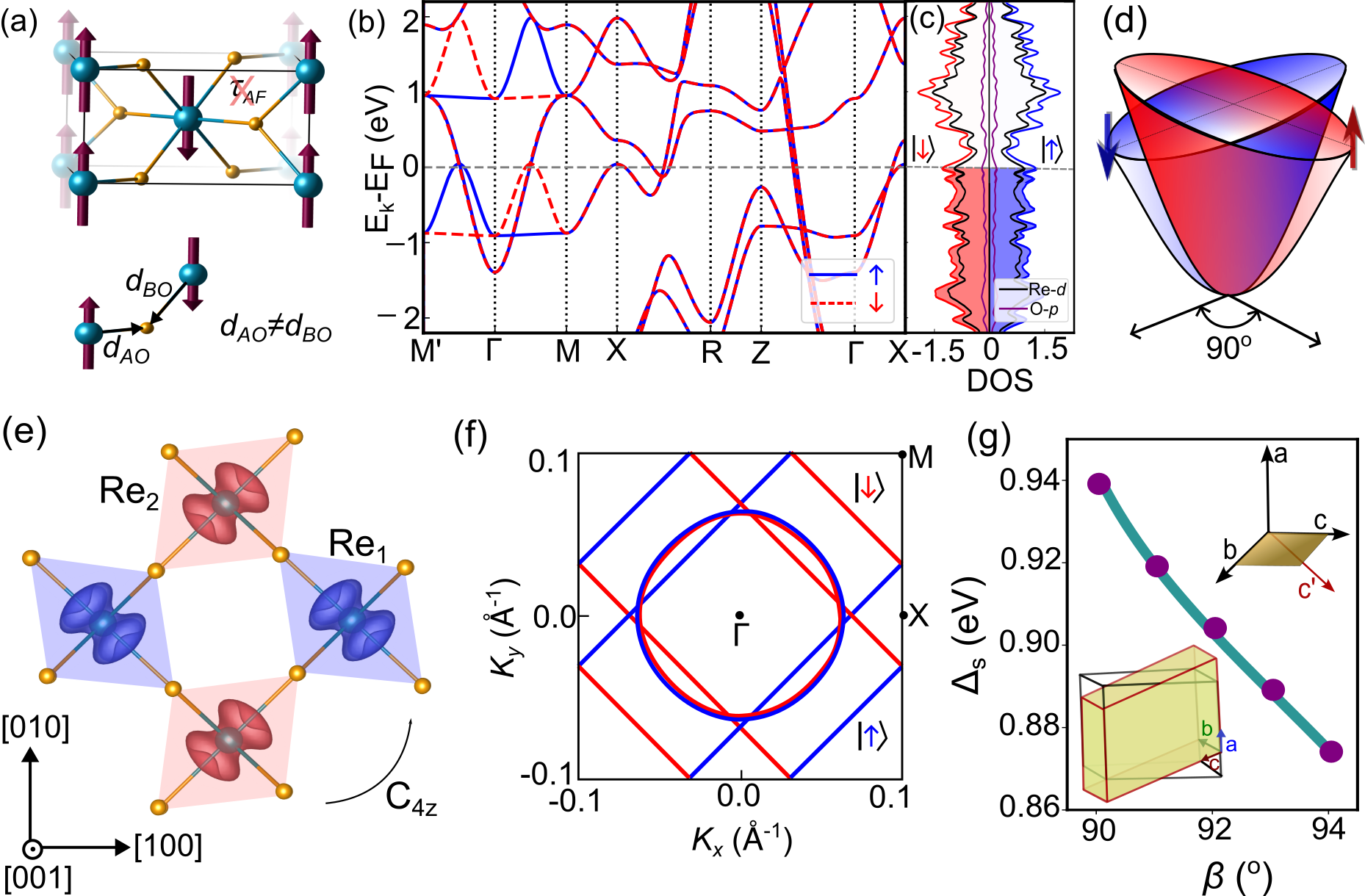}
 \caption{\textbf{Non-relativistic spin splitting:} (a) The crystal structure and bond connectivity of $R$-ReO$_2$ in ground state magnetic configuration. The inset shows the different Re-O bond lengths due to distribution of ligands over non-centrosymmetric sites.  (b) and (c) represents the spin resolved band and density of states plot (DFT) of $R$-ReO$_2$ in the absence of SOC. The energy axis is scaled w.r.t Fermi energy, $E_F$. The presence of a fully compensated density of states indicates zero magnetization, while the existence of momentum-dependent splitting in band dispersion indicates the broken TRS. (d) Schematics of momentum resolved spin splitting for $d$-wave AM candidate $R$-ReO$_2$. (e) Magnetization density plot of neighboring Re sub-lattices are orthogonal and preserves the four fold $C_{4z}$ rotational symmetry.  (f) Constant energy contour of  for $E=E_F$--0.15 eV.  This  shows the nature of the spin splitting is opposite for orthogonal momentum points.  (g) Change of spin splitting ($\Delta_s$) with deviation of crystallographic angle $\beta$ from $90^\circ$. The insets show the nature of deformation of the crystal unit cell with $\beta$ variation.
  \label{figure1}}
 \end{center}
\end{figure*}
%%%%%%%%%%%%%%%%%%%%%%%%%%%%%%%%%%%%%%%%%

In order to investigate the impact of pressure and to track the phase transition, we have carried out numerical calculations within DFT for different structures modulating the u.c. volume for both  phases. We have employed the Murnaghan equation of state~\cite{Murnaghan1944, Birch1947} to fit the obtained total energy  from self consistent DFT calculations:
\begin{equation}
    E(V)=E_0 +\frac{B_0 V}{B_0'} \Big [\frac{(V_0/V)^{B_0'}}{B_0'-1} +1 \Big ] -\frac{B_0V_0}{B_0'-1}.
\end{equation}
Here $V_0$ is the equilibrium volume and $B_0=-V(\partial P/\partial V)$ is the bulk modulus evaluated at $V_0$. The typical values of $B_0'$ (pressure derivative of $B_0$), used as a measure of stiffness of the material, are between 4 to 7~\cite{DeRaychaudhury2007}. We have included the variation of cohesive energy with volume in Fig.~\ref{fig_sch}~(c) for both $\alpha$- and $R$-ReO$_2$. It clearly shows that below a critical pressure, the compressive strain provides a way to stabilize the tetragonal structure w.r.t the usual monoclinic phase, in full agreement with the experimental observation~\cite{Shibata2020}. In the rest of the manuscript we establish that this is not merely a structural transition, rather it completely changes the nature of the magnetism promoting AFM (see Fig.~\ref{fig_sch}~(d)) to AM (see Fig.~\ref{fig_sch}~(e)) ground state in the same chemical composition of ReO$_2$. In the following we discuss the contrasting electronic, magnetic and topological properties for both $R$- and $\alpha$- ReO$_2$.

\subsection{Altermagnetic tetragonal $R$-ReO$_2$}
\subsubsection{Non-relativistic magnetic order and symmetry analysis}

Having established the structural transition mechanism to the high-pressure tetragonal phase, we now explore the electronic structure of the altermagnetic phase and characterize this from a microscopic and structural origin. The two Re sub-lattices of $R$-ReO$_2$, contained within the trigonally distorted octahedral environment, order with opposite spin moments, i.e. compensated magnetization (see Fig.~\ref{figure1}~(a)). To analyze the effect of crystal symmetry and the exchange-driven altermagnetic  ordering, we set initially the SOC to zero. Our DFT calculations within GGA+U scheme indicate that the state with anti-parallel arrangement of spins is substantially lower in energy of value 92.9~meV/ Re (111.8~meV /Re)  than its ferro-magnetic (non-magnetic) counterpart. Our result agrees well with the previous  first principle study on $R$-ReO$_2$~\cite{Ivanovskii2005}. The coexistence of metallic and AM behavior in $R$-ReO$_2$ is robust against variation of the Coulomb correlation (see section S2 of SM for details).  Although no experimental studies of tetragonal $R$-ReO$_2$ for magnetism have been reported, the presence of a AFM ground state has been proposed around $\sim$4.2K for its  orthorhombic $\beta$ phase~\cite{Wang2017}. Notably, the presence of anti-parallel spin ordering within the tetragonal stoichiometry has been studied and confirmed in the sister compounds e.g. RuO$_2$~\cite{Feng2022} and CoF$_2$~\cite{Disa2020}.

In Fig.~\ref{figure1}~(b) we plot the non-relativistic spin-polarized band structures along different high symmetry $k-$points for collinear compensated spin order. The total magnetization of the u.c. vanishes as can be seen in the density of states plot of Fig.~\ref{figure1}~(c). 
The distribution of ligand O-atoms over the non-centrosymmetric $4f$ Wyckoff positions promote unequal bond lengths with two different Re sub-lattices and 
give rise to the altermagnetic spin symmetry which interchanges sublattices with opposite spins via a rotation, and leads in turn to broken TRS in the electric band structure.
%triggers breakdown of effective TRS. 
We find momentum dependent splitting between up and down spins along the  $\Gamma  M$ path. The sign of the spin splitting is opposite for two orthogonal paths as can be seen between $\Gamma M$ and $\Gamma M^\prime$ directions in Fig.~\ref{figure1}~(a). The schematic of the spin splitting is plotted in Fig.~\ref{figure1}~(d) in agreement with its $d$-wave AM phase. The details of $d$-wave AM from the spin group theory perspective is included in section S1 of SM for completeness \cite{Smejkal2021a}. We have plotted the spin-densities of the two Re-sub-lattices in Fig.~\ref{figure1}~(e). The anisotropic magnetization densities of the two sub-lattices are connected through four-fold rotation crystal rotation. The constant energy contour near the Fermi energy, at $E_k-E_F$=-0.15 eV, shows that $\epsilon (\textbf{k}_\uparrow) \neq \epsilon(-\textbf{k}_\downarrow)$ confirming the break down of the TRS. Next we analyze our \textit{ab-initio} results from the crystal symmetry view point.

%%%%%%%%%% Figure 3 %%%%%%%%%%%%%%%%%%%%
\begin{figure*}[t!]
\begin{center}
  \includegraphics[width=1.85\columnwidth]{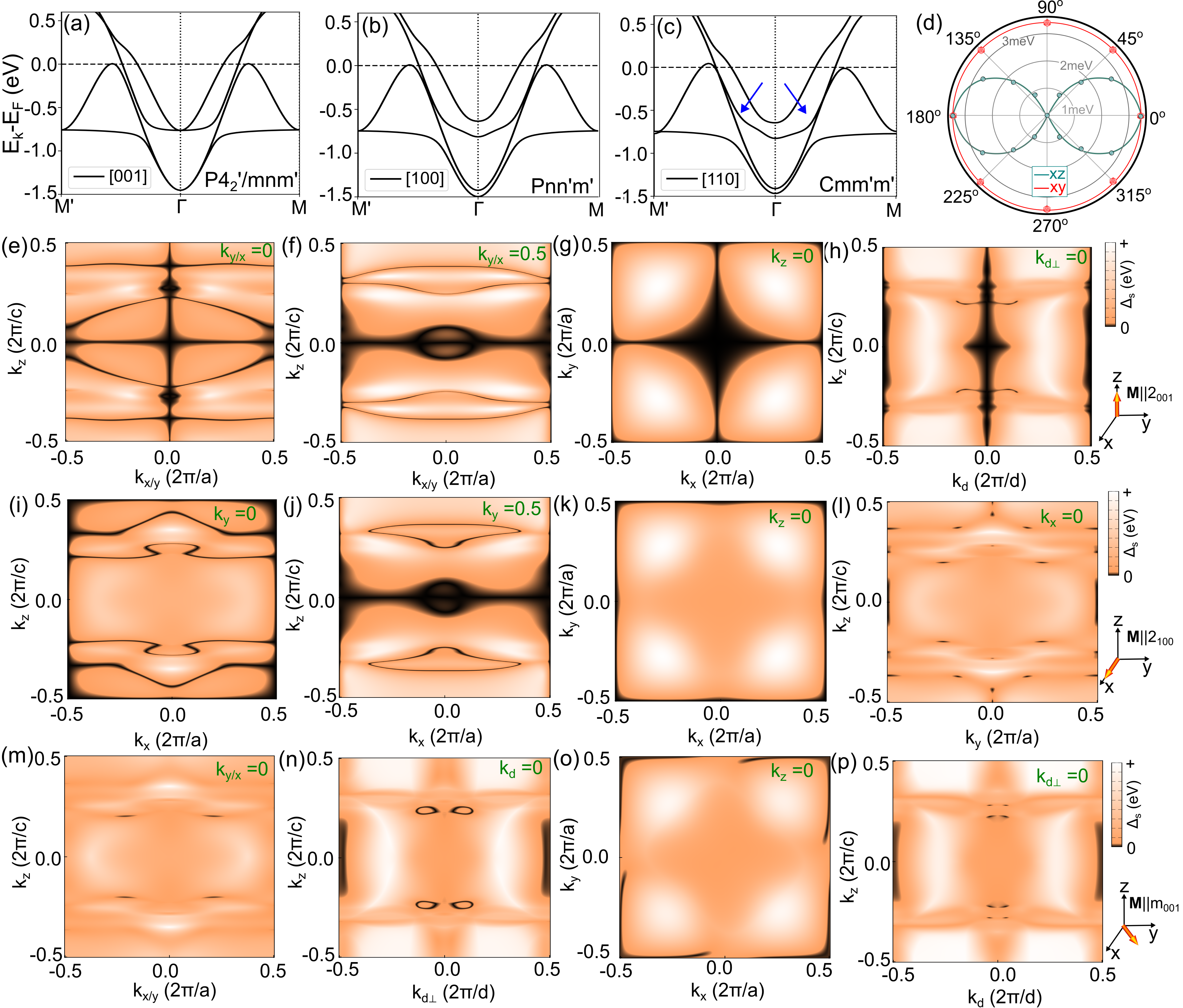}
  \caption{ \textbf{Spin-orbit coupling and Nodal crossing:} The band dispersion along two orthogonal spin splitted $\Gamma M$ high symmetry directions for out of plane [001] (a) and in plane spin quantization [100] (b) and [110] (c). The SOC induces asymmetry in the bulk band dispersion for magnetic anisotropy along [110]. (d) Calculated magnetic anisotropy energy (MAE) per unit cell when spin direction is rotated in the xz (green) and xy (red) planes in the AM configuration for U$_{eff}$ = 1.5 eV. The Re spins prefer easy axis magnetic anisotropy along $z$-axis. The nodal crossings between the first valence \textit{AP} for different planes are plotted for spin quantization along [001] (e-h), [100] (i-l) and [110] (m-p) respectively. Here $\hat{d}=\hat{x}+ \hat{y}$ and $\hat{d}_\perp=\hat{x}-\hat{y}$. The nodal crossings and Weyl points drastically changes for change in spin quantization. Due to the band asymmetry in (c), there exists nodal loop in (n) but no such nodal structures in (p).  \label{figure2}}
  \end{center}
\end{figure*}
%%%%%%%%%%%%%%%%%%%%%%%%%%%%%%%%%%%%%%%%%%
In the absence of SOC, the real space and the spin spaces are decoupled. The resulting non-relativistic space group has the following spin symmetry %representation %
\begin{equation}\label{eqn_ss}
    [E||mmm] + [C_2|| G\backslash mmm | \rm \bf{t}^\prime ],
\end{equation}
%[$E||mmm$]+[$C_2|| G\backslash mmm | t^\prime$], 
where $mmm$ and $G\backslash mmm$ include respectively \{$E$, $P$, $C_{2z}$, $C_{2d}$, $C_{2d \perp}$, $\mathcal{M}_z$, $\mathcal{M}_d$, $\mathcal{M}_{d\perp}\}$ and \{$C_{2y}$, $C_{2x}$, $\pm C_{4z}^{\pm}$, $\mathcal{M}_x$, $\mathcal{M}_y$\}. The $\hat{d}$ and $\hat{d}_\perp$ axes are along $\hat{x}+ \hat{y}$ and $\hat{x}-\hat{y}$ respectively. Here the operation of the left (right) of the parenthesis acts solely on the spin (real) space. The detailed description of the symmetry operations are included in section S1 of SM. Here $\textbf{t}^\prime$ is the half translation along the body-diagonal. If the opposite spin sublattices are not connected by a translation or inversion, the Kramer's degeneracy of the spin up and down channels get lifted, promoting an alternating Zeeman like band splitting even in the absence of SOC, and connecting opposite spin states  whose momentum is related by the $C_{4z}$ rotation. 
We refer these opposite  spin bands as altermagnetic-pair (\textit{AP}) throughout the rest of the article. The two sub-lattices containing opposite spins are related by the screw rotation \{$C_{4z}|\textbf{t}^\prime$\} in $R$-ReO$_2$, as can be seen in the relative octahedral orientation and spin density plot of Fig.~\ref{figure1}~(e). %Here translation $\textbf{t}^\prime=(1/2, 1/2, 1/2)$. 
As a result, we see a spin splitting between \textit{AP} bands along the $\Gamma M$ paths within the non-relativistic limit. As demonstrated in Eq.~\ref{eqn_ss}, the $[C_2||\mathcal{M}_{x(y)}]$ yields $\epsilon(k_x, k_y,k_z,\sigma) =\epsilon(\mp k_x, \pm k_y, k_z, -\sigma) $. Thus, the opposite spin \textit{AP} bands are degenerate for both $k_x=0, \pi$ and $k_y=0, \pi$ planes within the non-relativistic limit as shown in Fig. S3 of section S4 of SM.

Next we discuss the impact a small structural deformation has on the non-relativistic energetics. The crystal symmetry analysis suggest that the primary difference between the monoclinic u.c. and its tetragonal u.c., is the deviation of the crystallographic angle $\beta$ from $90^\circ$, i.e. the crystal axis $c$ is no longer perpendicular to the $a-b$ plane, as shown in the inset of Fig.~\ref{figure1}~(g). To induce this typical distortion we have constructed hypothetical structures from the $R$-ReO$_2$ u.c. by manually increasing the angle $\beta$ slightly from $90^\circ$. For energy minimization, we have done ionic relaxation of each structure using the conjugate-gradient algorithm until the Hellman-Feynman forces on each atom are less than the tolerance value of 0.01 eV$\slash {\rm \AA}$. In Fig.~\ref{figure1}~(g) we show the change in spin splitting with angle $\beta$. The reduction in spin splitting with increase in $\beta$ drives the system towards an AFM state. However, there exist other effects in realistic situations, such as various magnetic arrangements, atomic reconstructions, difference between magnetic and crystal unit cell (i.e. $\textbf{Q}\ne0$), etc. Hence the details of electronic structure in the monoclinic phase is discussed later in section-\ref{sec_monoclinic}. \\

\subsubsection{Effect of spin-orbit coupling and magnetic anisotropy}
The presence of heavy $4d/5d$ transition metal elements within any system can %exert a profound 
influence the strength of the SOC, ultimately determining the ground state characteristics of the magnetic order~\cite{Kumar2019, Chakraborty2019, Chakraborty2021, Bhowal2021,Bandyopadhyay2022}. In the following we investigate how the crystal symmetry analysis of the spin splitting proceeds in the relativistic regime. 
We next incorporate SOC in our calculations, leading to magnetic anisotropy that leads to a preferred axis quantization. This leads to different magnetic space groups for the different quantization axis, relevant for analyzing the topological properties of their respective states.

In Fig.~\ref{figure2}~(a),~(b) and (c) we plot the relativistic band dispersion with spin along [001], [100] and [110] axes respectively. The corresponding magnetic space groups are marked in the respective plots. As the magnetic Re atoms occupy the Wyckoff site 2a in the unit cell, the ground state magnetic space group only allows collinear magnetic arrangements (see allowed moment per Wyckoff positions for ground state configuration in Table-I of section S4 of SM even in the presence of SOC. We find that changing the spin quantization from out-of-plane to in-plane direction now opens up the degeneracy of the opposite spin bands at the $\Gamma$ point, i.e. a small net magnetization is present, proportional to the SOC. Our $ab-initio$ calculations predict a  [001] spin quantization axis with magneto-crystalline anisotropy (MCA) energy $\sim$3 meV. The polar plot of the MCA energy in shown in Fig.~\ref{figure2}~(d) when the spin directions are rotated on the $xz$ (green) and $xy$ (red) planes respectively. This low anisotropy energy makes $R$-ReO$_2$ a promising material for adjusting the N{\'e}el vector for anomalous transport experiments. Our calculation suggests that finite electron doping can promote in-plane magneto-crystalline anisotropy in $R$-ReO$_2$ (see section S3 of SI for detailed discussion). The ground state characteristics of $R$-ReO$_2$ closely resemble those of the significant $d$-wave AM  RuO$_2$.

%%%%%%%%%%%%%Figure 4%%%%%%%%%%%%%%%%%%%%%%%%%%
\begin{figure*}[t!]
\begin{center}
  \includegraphics[width=2\columnwidth]{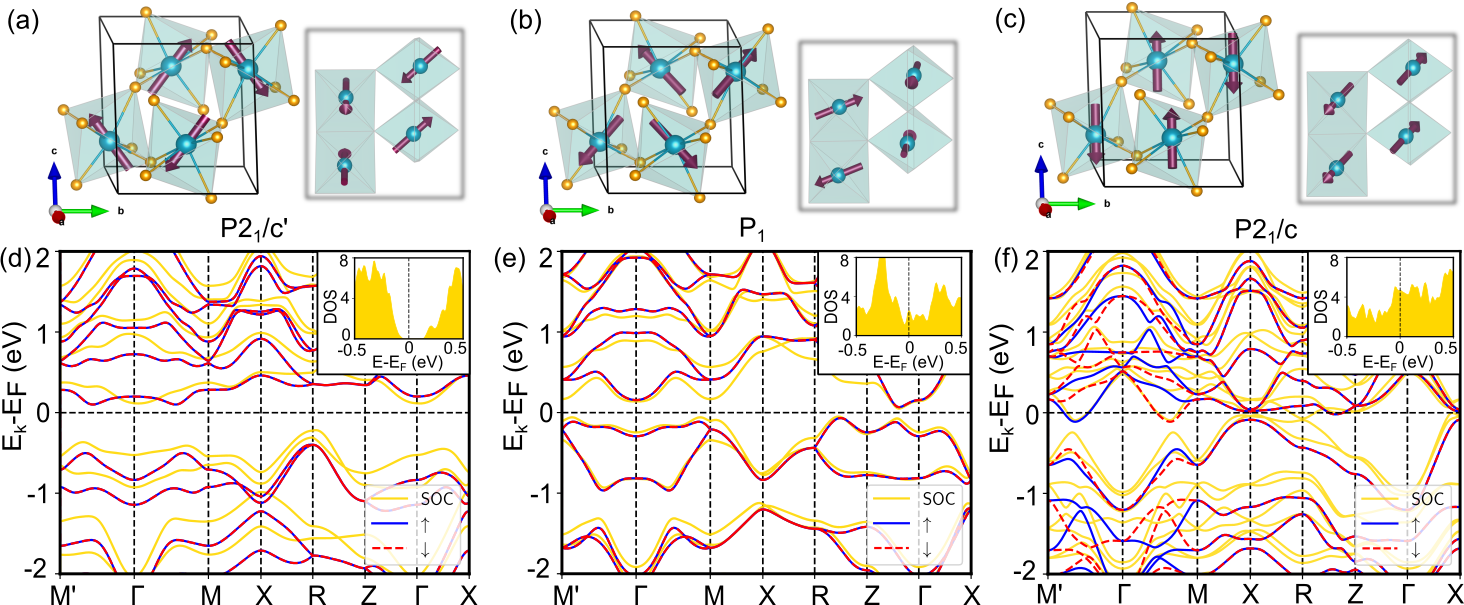}
  \caption{ \textbf{Monoclinic magnetic states:} Three possible anti-parallel arrangement of spin of monoclinic $\alpha$ in (a), (b) and (c) respectively. Spin polarized band dispersion both without (blue solid and red dashed lines) and with SOC (yellow solid line) are plotted in (d), (e) and (f) for spin configurations shown in (a), (b) and (c) respectively. Insets of (d-e) show corresponding density of states for each magnetic of (a-c) configurations. The magnetic configuration plotted in (a) is the lowest energy state of $\alpha$-ReO$_2$. 
  \label{figure4}}
  \end{center}
\end{figure*}
%%%%%%%%%%%%%%%%%%%%%%%%%%%%%%%%%%%%%%%%%%%%%%%
%\subsubsection{Anti-Krammer's splitting and Nodal crossings}
\subsubsection{Relativistic splitting and Nodal crossings}
In this section we will discuss how inclusion of SOC additionally splits the \textit{AP} in certain high symmetry directions and changes the nodal structures. Hence the nodal planes /lines/ points present in the non-relativistic regime are no longer preserved or break into parts depending on the spin quantization. First we analyze the possibility of a nodal plane for the ground state magnetic configuration, with
quantization axis [001] and magnetic space group $P4_2'/mnm'$, containing the following symmetry operations:

\begin{widetext}
\begin{center}
$ \{ E, P, C_{2z}, \mathcal{M}_z\}~+~\textbf{t}^\prime \{ C_{2x}, C_{2y}, \mathcal{M}_x, \mathcal{M}_y \}~+ \mathcal{T} \{ C_{2d}, C_{2d\perp}, \mathcal{M}_d, \mathcal{M}_{d\perp} \} + \textbf{t}^\prime \mathcal{T} \{\pm C_{4z}^{\pm}\}$.  
    
\end{center} 

\end{widetext}    
%It is evident from the above symmetry analysis that the nodal plane can be only protected in the $k_z=\pm \pi$ plane, 
Here $\mathcal{T}$ is the time-reversal operation. $\mathcal{M}_d$ and $\mathcal{M}_{d\perp}$ mirror planes, connecting the opposite sublattices, can host the nodal crossings. 
The non-relativistic nodal crossings which are preserved in the presence of SOC, are therefore those that intersect with these $\mathcal{M}_d$ and $\mathcal{M}_{d\perp}$ mirror planes, i.e. $\Gamma Z$, $AM$ and $AM'$ lines.  
%Hence only those non-relativistic nodal crossings are preserved which is at the intersection of $k_{x,y}=0, \pm \pi$ intersects the relativistic mirror planes. 
These sharp nodal lines in presence of SOC can be seen in Fig.~\ref{figure2} (e).

Next we explore the possibility of any other nodal planes within the ground state $R$-ReO$_2$. The operation $\textbf{t}^\prime \mathcal{T} \{\pm C_{4z}^{\pm}\} = \mathcal{T} S_{4z}^\pm \equiv \mathcal{A}$, %\textcolor{blue}{[alternatively $[C_2||(\pm C_{4z}^{\pm}) |t^\prime] = [C_2||S_{4z}^{\pm}]$]} 
where the combined effect of space operation and time reversal maps the spins of the two sub-lattices. The Hamiltonian is invariant under the symmetry operation $\mathcal{A}$ i.e. $\mathcal{A}^{-1}\mathcal{H}\mathcal{A}=~\mathcal{H}$. Below we analyze the symmetry protection of the nodal planes in the presence of SOC.
We consider two block wave-functions $\psi_{\bf k}$ and  $\psi^\prime_{\bf k}$  which are related by this non-symmorphic operation as $\psi^\prime_{\bf k^\prime}=\mathcal{T} S_{4z} \psi_{\bf k}$. The eigenvalue equation for momentum ${\bf k}^\prime$, where ${\bf k}^\prime=\mathcal{A}{\bf k} \neq {\bf k}$:
\begin{equation}
E_{k^\prime}\psi^{\prime}_{{\bf k}^\prime} = \mathcal{H}\psi^{\prime}_{{\bf k}^\prime} = \mathcal{H}[\mathcal{A} \psi_{\bf k}], 
\end{equation} 
\begin{equation}
\mathcal{A}[\mathcal{H}\psi_{\bf k}]= E_{\bf k} [\mathcal{A}\psi_{\bf k}].
\end{equation}

As the Hamiltonian is invariant under the symmetry operation connecting the two eigenfunctions, these must share common eigenvalues. Now to check the orthogonality condition we calculate the overlap function,
\begin{equation}\label{eq_overlap}
    <\psi_{\bf k}|\psi^{\prime}_{{\bf k}^\prime}> = <\mathcal{A}\psi^{\prime}_{{\bf k}^\prime}|\mathcal{A}\psi_{\bf k}> = \mathcal{A}^2 <\psi^{\prime}_{{\bf k}^\prime}| \psi_{\bf k}> 
    %e^{i(k_y-k_x+k_z)/2} e^{i(k_x+k_y+k_z)/2}= e^{i(k_y+k_z)}.
\end{equation}
To satisfy the orthogonality condition, the L.H.S and R.H.S of Eq.~\ref{eq_overlap} must be opposite in sign.  By definition of the space operation, $\mathcal{A}^2=e^{i(k_y-k_x+k_z)/2} e^{i(k_x+k_y+k_z)/2}= e^{i(k_y+k_z)}$. The two wave-functions are orthogonal  only if the wave vector satisfies $k_y+k_z=\pi$. But, non-relativistic nodal planes exist only for planes $k_x= 0, \pm \pi$ and $k_y= 0, \pm \pi$ (see Fig. S3 in section S4 of SM) which does not satisfy the criteria of Eq.~\ref{eq_overlap}. Therefore for the ground state $P4_2'/mnm'$, we don't have any nodal planes. Instead we find nodal lines and Weyl points as depicted in Fig.~\ref{figure2}.

%%%%%%%%%%%%%%% Figure 5 %%%%%%%%%%%%%%%%%%%
\begin{figure*}[t!]
\begin{center}
\includegraphics[width=1.9\columnwidth]{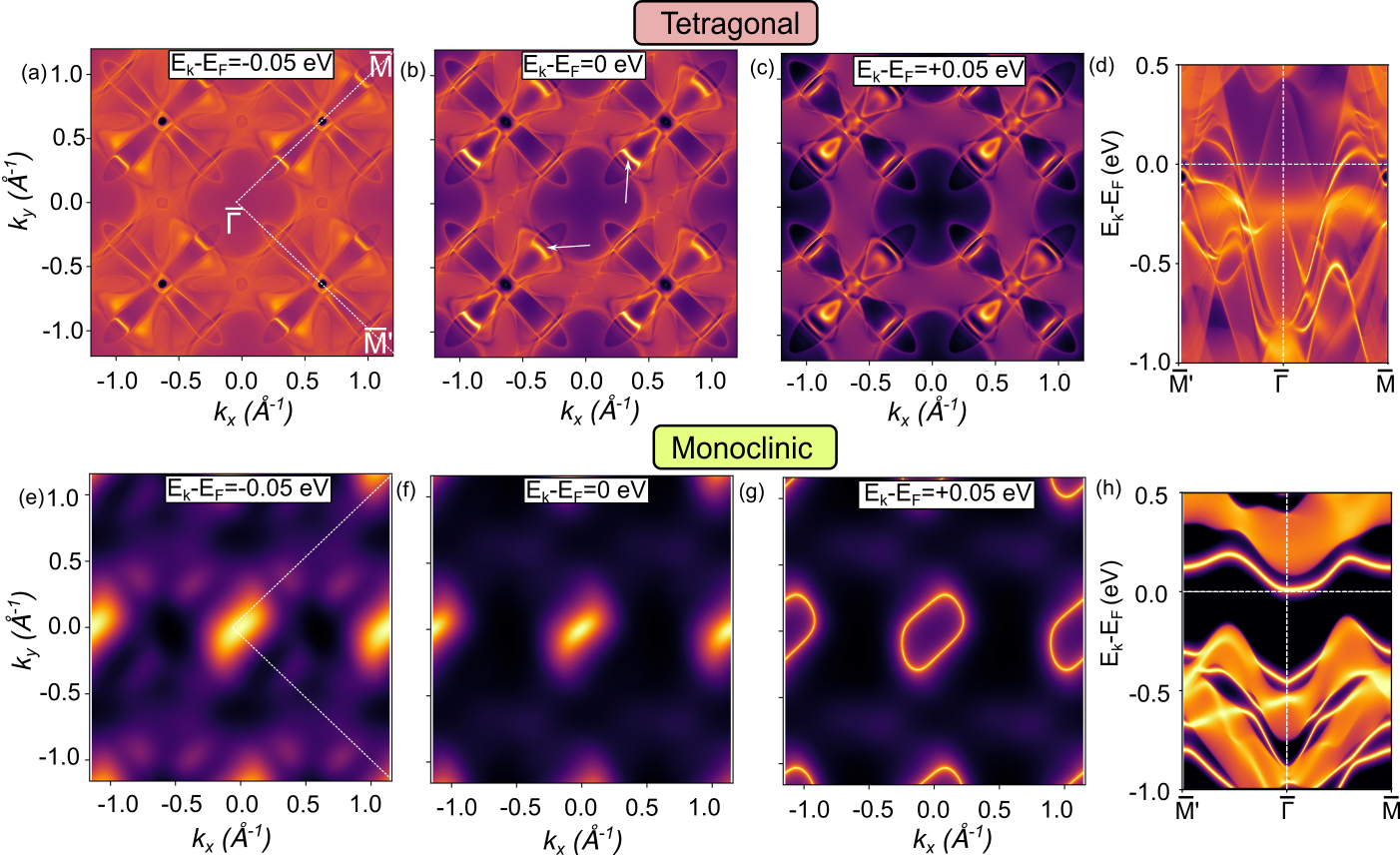} 
\caption{\textbf{Spectral function and surface state topology:} The surface state arc near the Fermi energy for both tetragonal $R$-ReO$_2$ (a-c) and monoclinic $\alpha$-ReO$_2$ (e-g) phases. The spectral functions over two orthogonal $\Gamma M$ momentum paths are shown as a function of energy in (d) and (h) for tetragonal and monoclinic phases respectively. The open surface Fermi arcs along the $\Gamma M$ directions of $R-$ReO$_2$ are shown with white arrows. (a) (c) shows the arc structure at $E_F$ (b) is robust against small change in the energy may originate from finite surface potentials.  We observe closed loop surface spectral function slightly below the Fermi energy, at E=E+0.05 eV for monolcinic plase as shown in (f). The surface state get detach with increasing energy value as seen in (h). \label{figure5}}
  \end{center}
\end{figure*}

The band crossings between \textit{AP} lead to several nodal lines and Weyl points. Different crystallographic planes containing the nodal crossings between the lowest valence AP for spin quantization along [001] are plotted in Fig.~\ref{figure2}~(e-h). Due to the equivalence of $k_x$ and $k_y$ axes relative to spin quantization direction, swapping between $k_x$ and $k_y$ in momentum space leads to an identical nodal structure, as shown in Fig.~\ref{figure2}~(e) and (f). 
Further, in Fig.~\ref{figure2} (i-l) and (m-p), the nodal crossings are plotted for  magnetic space groups  $Pnn'm$, and $Cmm'm'$, where we have easy axis spin anisotropy along [100] for the former and easy plane [110] spin quantization for the later respectively. 
The nodal lines on the mirror planes will be robust against infinitesimal field perpendicular to the mirror planes. The nodal lines of \textit{AP} bands will be protected when the opposite spins directions are parallel (perpendicular) to the mirror plane $\mathcal{M}$ ($\mathcal{M}^\prime\equiv\mathcal{T} \mathcal{M})$. The SOC induces a band asymmetry along orthogonal $\Gamma M$ directions, as can be seen in 
Fig.~\ref{figure2} (c). As a result we get nodal loops in the ${\bf k}_{110}\equiv {\bf k}_d$ plane whereas the orthogonal plane ${\bf k}_{1\bar{1}0} \equiv {\bf k}_{d\perp}$ does not contain that. Hence, the SOC not only settles the magnetic anisotropy energy for feasibility in anomalous transport applications but also dictates the nature of non-trivial nodal lines.

\subsection{Magnetic ground state of monoclinic $\alpha$-ReO$_2$} \label{sec_monoclinic}
In this section we will discuss the nature of magnetism for monoclinic phase. The different arrangement of Re magnetic moments over four possible $1a$ Wyckoff sites can lead to reduction in crystal symmetries and consequently affect the electronic structure and topology of the system. The nearest neighbor ReO$_6$ octahedra form edge shared linear network in [$\bar{1}$,1,$\bar{1}$] direction. The two nearest edge shared octahedral chains are  connected by corner sharing. The allowed anti-parallel arrangements of moments at $q=0$ can be two fold: A. anti-ferromagnetic ordering along edge shared and combination of anti-ferromagnetic and ferromagnetic ordering between half of the corner shared octahedra. B. anti-ferromagnetic (ferromagnetic) ordering along all corner (edge) shared octahedra.  The two Fig.~\ref{figure4}~(a) and (b) belongs to former and later category of kind-A. Here the two edge-shared sub-lattices are simply connected by a inversion combined with time reversal $\mathcal{T}$. As a result within the non-relativistic limit the up and down spin channels are completely degenerate (see Fig~\ref{figure4} (d), (e)) through out the whole BZ. Inclusion of SOC stabilizes a relative spin canting between the corner shared sites for both Fig~\ref{figure4} (a) and (b) magnetic arrangements. There exist two stable compensated non-collinear magnetic arrangements originating from Kind-A belonging to the different magnetic space groups $P2_1/c^\prime$ and $P2_{1^\prime}/c$ due to different spin quantization axis. The edge shared spins are rotated spatially by 68.8$^\circ$ and 88.2$^\circ$ for $P2_1/c^\prime$ and $P2_{1^\prime}/c$ respectively. Our theoretical calculation within GGA+SOC+U scheme suggest the later to be 17.3 meV/u.c. higher in energy than the AFM ground state $P2_1/c^\prime$. The magnetic space group symmetry of the ground state $P2_1/c^\prime$ includes $\epsilon$, $P\mathcal{T}$, $\mathbf{t}_m C_{2y}$ and  $\mathbf{t}_m\mathcal{T} \mathcal{M}_y$; where translation vector $\mathbf{t}_m$ is ($0,\frac{1}{2},\frac{1}{2}$). The SOC lifts the four fold degeneracy of the bands, but Kramer's degeneracy remain intact even in the presence of SOC due to its $P\mathcal{T}$ symmetry. We have plotted the spin configuration in Fig.~\ref{figure4} (c) and corresponding band dispersion (in Fig.~\ref{figure4} (f)) for higher energy state of kind-B. Here the Re sites with opposite spins are only connected through $\mathcal{T} \{C_4|t\}$ which offers a possibility of \textit{AP} bands from symmetry consideration. The splitting between \textit{AP} is relatively small ($\sim$ 0.3 eV) which is of the order of SOC induced band splitting of the system [see green line in Fig~\ref{figure4} (f)]. The AM $P2_1/c$ configuration and FM state is 28.6 meV and 185.9 meV  higher than the AFM ground state of the monoclinic phase. Among all these possible magnetic orders, AFM spin configuration commensurate with magnetic space group $P2_1/c^\prime$ has the lowest energy. Contrary to metallic AM state of tetragonal phase, the ground state of monoclinic $\alpha$-ReO$_2$ is insulating with band gap value of $\sim$0.19 eV. Along with strain induced AFM to AM transition from $\alpha$- and $R$- ReO$_2$, it also undergoes MIT which is a useful experimental signature to track the phase transition. 

%%%%%%%%%%%%%%%%%%%%%%% Figure 6 %%%%%%%%%%%%%%%%%%%%%%
\begin{figure*}[t!]
\begin{center}
\includegraphics[width=2\columnwidth]{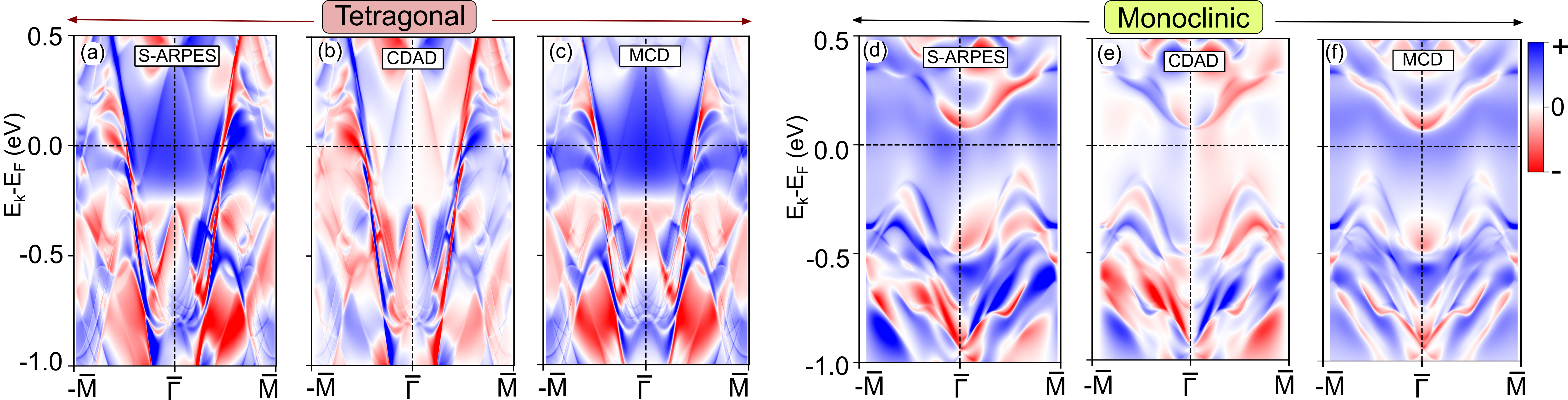} 
\caption{\textbf{Spin-ARPES calculations:} Spin polarized surface spectral function, $S(\textbf{k}_{||},\omega)$, for both $R$-ReO$_2$ (a) and $\alpha$-ReO$_2$ (d) along two opposite momentum paths.  the anti-symmetric part, CDAD (symmetric part. MCD) of the S-ARPES for opposite momentum transfer in plotted in b (c) and e (f) for $R$-ReO$_2$ and $\alpha$-ReO$_2$ phases respectively. The MCD spectra of tetragonal phase is primarily contributed from the bulk states conveying the internal TRS breaking of the AM $R$-ReO$_2$. Both of the CDAD and MCD in $\alpha$-ReO$_2$ originate the broken $P$ and $\mathcal{T}$ symmetry, combined with SOC.
\label{figure6}}
\end{center}
\end{figure*}
%%%%%%%%%%%%%%%%%%%%%%%%%%%%%%%%%%%%%%%

%%%%%%%%%%%%%%%
\subsection{Contrasting %non-trivial 
surface states between AM and AFM phases}
Conventionally the $R$- and $\alpha$- ReO$_2$ belongs to two different elementary classes of magnetism. The $P\mathcal{T}$ is preserved for AFM ground state of $\alpha$-ReO$_2$,  whereas the opposite sublattices in $R$-ReO$_2$ are  connected by the $C_{4z}\mathcal{T}$ symmetry.  As a result we find strong momentum dependent \textit{AP} splitting of $\sim$1 eV in $R$-ReO$_2$ while Kramer's degeneracy is present for the monoclinic ground state.  These dissimilarities in magnetic nature percolates to the sharply contrasting surface properties of these phases, which can be observed by, e.g., angle resolved photo-emission spectroscopy (ARPES). The surface spectral function can be obtained from the following relation of retarded Green's function ($G_s$)~\cite{WU2018, Damascelli2004}:
\begin{equation}
    A(\textbf{k}_{||}, \omega)= -\frac{1}{\pi} \lim_{\eta\to 0^+} \mathrm{Im}~\mathrm{Tr} [G_s (\textbf{k}_{||}, \omega+ i\eta)] 
\end{equation}\label{arpes_eqn}

Fig.~\ref{figure5} includes the constant energy contours calculated by creating a finite slab geometry along the (001) direction with magnetic anisotropy along z-axis.  The arc of spectral functions of $R$- ($\alpha$-) ReO$_2$ are plotted for $E-E_F=$ -0.05 eV, 0~eV and +0.05~eV in a(e), b(f) and c(g) respectively. There exist pair of open Fermi surface arcs along $\bar{\Gamma}\bar{M}$ directions. Given the surface sensitivity and probing depth of the experimental technique, there exist a finite uncertainty of the Fermi energy. We find that these paired surface arcs are robust with small variation of energy around $E_F$ that can be induced by surface potentials. In Fig.~\ref{figure5}~(f) we see the Fermi arc forms a closed loop structure around the $\bar{\Gamma}$ point near $E_F$ for the monoclinic phase. The surface loop reside well within the bulk gap without connecting the conduction and valence sectors which makes it trivial in nature. The distorted surface loop shows the asymmetric nature along two orthogonal directions of the BZ. With decreasing binding energy, the surface arc detaches from the Fermi energy as can be seen in Fig.~\ref{figure5}~(e), (f).

In Fig.~\ref{figure5}~(d) and (h) the spectral function contributed from both bulk and surface states are plotted as a function of energy along orthogonal $\bar{M}^\prime\bar{\Gamma}\bar{M}$ paths (white dashed line in Fig.~\ref{figure5}~(a)) for tetragonal and monoclinic phases respectively. We see an isolated surface state within the bulk gap just above the Fermi energy of $\alpha$-ReO$_2$. The asymmetry of the surface state between two orthogonal paths verifies the distorted surface loop structure of Fig.~\ref{figure5}~(g).

\subsection{Spin-polarized spectral functions}
There exists unique spin textures in the momentum resolved surface spectral function of altermagnetic and antiferromagnetic materials as a consequence of the nontrivial ground state magnetism. In this section we compare the spin polarized ARPES spectra between the two phases. The spin polarized spectral functions ($S(\textbf{k}_{||}, \omega)$) can be calculated from the following expression:
\begin{equation}
    S(\textbf{k}_{||}, \omega)=-\frac{1}{\pi} \lim_{\eta\to 0^+} \mathrm{Im}~\mathrm{Tr} [ \bm{\sigma} G_s (\textbf{k}_{||}, \omega+ i\eta)] / A(\textbf{k}_{||},\omega). 
\end{equation} 
We have plotted the spin polarized spectral functions along two opposite momentum $\bar{\Gamma}\bar{M}$ paths for both $R$- and $\alpha$-ReO$_2$ in Fig.~\ref{figure6}. The pronounced asymmetric nature of S-ARPES confirms the presence of both symmetric and anti-symmetric part with respect to the opposite momentum direction. First we consider the spectral function plots of tetragonal phase as shown in Fig.~\ref{figure6}~(a-c). We  calculate circular dichroism in the angular distribution (CDAD),  arising from the experimental geometry. CDAD is primarily anti-symmetric for opposite momentum due to broken inversion symmetry at the surface. Apart from the CDAD, there exists additional symmetric part of the Spin-polarized ARPES (S-ARPES), namely magnetic circular dichroism (MCD), which indicates the spin-polarization of the system and is directly connected to TRS braking in momentum space. Comparison of Fig.~\ref{figure6}~(b) and (c) with spin integrated spectral function of along $\bar{\Gamma}\bar{M}$ path (see in Fig.~\ref{figure5}~(d)), suggests the CDAD is pronounced in the surface state dominated momentum points. Contrary the MCD arises primarily from bulk spectral regions for $R$-ReO$_2$. { The significantly larger magnitude of the MCD spectra contributed from bulk states substantiate the intrinsic TRS breaking within the bulk beyond any surface dominated affects, whereas CDAD counterpart comes primarily due to surface geometry, near the Fermi energy.} Recent experimental studies report symmetric MCD spectra as an identification of the TRS breaking spin splitting in sister altermagnetic tetragonal compound RuO$_2$~\cite{Fedchenko2023, Krempasky2023, Lee2024}.
%%%%%%%%%%%%%%%%%%%%%%%%%%%%%%%%%%%%%%%%%%%%%%

Next we  analyze the $\alpha$-ReO$_2$ S-ARPES plots for the (001) surface convoluted with the bulk in Fig.~\ref{figure6}~(d-f). Unlike the tetragonal phase, both CDAD and MCD have comparable intensity. The absence of individual $P$ and $\mathcal{T}$ symmetry within the ground state AFM configuration {coupled with surface mediated contributions} lead to asymmetry between the states with $E_{\mathbf{k}}$ and $E_{\mathbf{-k}}$, which essentially gives rise to different spin-ARPES spectra for opposite momentum~\cite{Fedchenko2022}. This asymmetry within the spin-polarized spectral function of the ground state AFM leads to small MCD contribution at the $\alpha$-ReO$_2$. The calculation of these spectral features as a function of strain, demonstrate in particular how they arise from different physical origin.

\section{Conclusion}
In summary we have presented strain induced structural and magnetic transition accompanied with MIT between conventional $P\mathcal{T}$ symmetric AFM to AM phase for exploring technical applications. The bulk monoclinic $\alpha$-ReO$_2$ phase is compensated AFM in nature with finite spin canting between two neighboring corner shared octahedral Re sites.
The ground state of the monoclinic phase is an insulator  and accomodates Kramer's degenerate doublets, even in presence of SOC. Whereas, high pressure $R$-ReO$_2$ phase hosts unique altermagnetism with $d-$wave spin order. The distinct magnetic order between the two phases reflects in the contrasting surface states spectra. The pronounced MCD spectra obtained from our spin-polarized spectral function calculations confirms the intrinsic exchange driven TRS broken state in $R$-ReO$_2$, which serves as an experimental verification of the AM state. %The two phases connect through continuous shear strain very similar to the martensitic structural transition of non-magnetic ZrO$_2$ systems. 
Our study provides the first evidence of strain-tuning conventional AFM state to the new third category AM magnets. 

\section{Methods}
To perform the {\it ab-initio} calculations, we used density functional theory (DFT) in the plane wave basis set. We used the Perdew-Burke-Ernzerhof (PBE)~\cite{Perdew1997} implementation of the generalized gradient approximation (GGA)  for the exchange-correlation. This was combined with the projector augmented wave potentials~\cite{Blochl1994, Kresse1999} 
as implemented in the Vienna {\it ab initio} simulation package (VASP)~\cite{Kresse1993, Kresse1996a}. GGA calculations are carried out with and without Coulomb correlation (Hubbard U) and spin-orbit coupling (SOC). The calculations were done with the usual value of $U$ and Hund’s coupling $J_H$ chosen for Re with $U_{\mathrm{eff}} (\equiv U-J_H)$=1.5 eV in the Dudarev scheme~\cite{Dudarev1998}. The SOC is included in the calculations as a second variational form to the original Hamiltonian. The kinetic energy cutoff of the plane wave basis for the DFT calculations was chosen to be 520 eV. A  $\Gamma$-centered $9\times9\times 17$ and $10\times10\times 10$ $k$-point grids are used to perform the momentum-space calculations for the  Brillouin zone (BZ) integration of tetragonal and monoclinic phases respectively. To calculate the surface spectral function for finite geometry slabs of $R$- and $\alpha$- ReO$_2$, we construct the tight-binding model Hamiltonian by using atom-centered Wannier functions within the  VASP2WANNIER90 \cite{Souza2001} codes. 
Utilizing the obtained tight-binding model, 
we calculate the surface spectral function using the iterative Green’s function method, as implemented in the WannierTools package~\cite{Wu2017b}. 

\section{Acknowledgment}
AC acknowledges financial support from Alexander von Humboldt postdoctoral fellowship. We acknowledge funding by the Deutsche Forschungsgemeinschaft (DFG, German Research Foundation)-TRR288-422213477 (project A09). LS acknowledges support by Grant Agency of the Czech Republic grant no. 19-28375X. We acknowledge high performance computational facility of supercomputer `Mogon' at Johannes Gutenberg Universit\"{a}t Mainz, Germany. We acknowledge Roser Valenti and Rodrigo Jaeschke for stimulating discussions. \\

\section{Author Contribution}
A.C. carried out the theoretical calculations and co-wrote the manuscript. J.S. supervised the project and co-wrote the manuscript. R.G.H and L.S. participated in the discussions and provided useful feedback. All authors discussed the results and revised the manuscript.

\clearpage
%\bibliography{Complete_reference}
%#################################################%

%
\end{document}